\documentstyle[aps,preprint]{revtex}
\tightenlines
\begin{document}
\draft
\title{Statistics Of The Burst Model At Super-critical Phase}
\author{T. C. Chan\footnotemark[1]\footnotemark[2], H. F.
 Chau\footnotemark[3]\footnotemark[4], and K. S. Cheng\footnotemark[1]}
\address{
\footnotemark[1]Department of Physics, University of Hong Kong, Pokfulam Road,
 Hong Kong\\
\footnotemark[3]School of Natural Sciences, Institute for Advanced Study, Olden
 Lane,\\ Princeton, NJ 08540, USA
}
\preprint{IASSNS-HEP-95/62}
\date{\today}
\footnotetext[2]{Present address: Department of Physics and Astronomy, Rutgers
 University}
\footnotetext[4]{Corresponding author, e-mail: chau@sns.ias.edu}
\maketitle
\mediumtext
\begin{abstract}
We investigate the statistics of a model of type-I X-ray burst [Phys. Rev. E,
{\bf 51}, 3045 (1995)] in its super-critical phase.  The time evolution of the
burnable clusters, places where fire can pass through, is studied using simple
statistical arguments.  We offer a simple picture for the time evolution of the
percentage of space covered by burnable clusters.  A relation between the
time-average and the peak percentage of space covered by burnable clusters is
also derived.
\end{abstract}
\medskip
\pacs{PACS number: 05.70.Ln, 05.40.+j, 64.60.-i, 64.80,-v}
\widetext
\section{Introduction}
In recent years, spatially extended systems of many degrees of freedom have
attracted considerable attention.  A number of cellular automaton models,
for example, Eden model \cite{Eden}, percolation \cite{Perc}, contact
process \cite{CP}, and various self-organized critical models \cite{SOC,FFM}
have been investigated.  The forest fire model (FFM) \cite{FFM} and some
sandpile models with special geometries or toppling rules \cite{SP} tell us
that the existence of local particle conservation law is not a necessary
condition for self-organized critical (SOC).  Thus, some of the SOC models
may be used to comprehend the underlying principles of dissipative systems.

As non-conserving systems are divers in many branches of study, we have
proposed a simple cellular automaton model for diffusive and dissipative
systems \cite{Chan95}, whose rules are similar to that of the FFM.  Our model
is applicable to a number of phenomena including the type-I X-ray bursts
\cite{x-ray}, and the CO$_2$ gas outburst in some crater lakes \cite{lake}.
Because of this, we name our model the ``X-ray burst model'', or simply the
``burst model''. We studied the burst model by both numerical and
semi-analytical means and we found that the model exhibits various phases
when choosing different set of parameters \cite{Chan95,x-ray}. In spite of the
similarity with the FFM, the burst model is not self-organized critical.

In this paper, we focus on the statistical behavior of the burst model at its
super-critical phase.  First, we briefly review the burst model in Section~II.
Then we discuss the relationship of the burst model at its super-critical phase
with percolation in Section~III.  A relation between the time-average coverage
fraction and the average coverage fraction just before burning (precise
definitions will be given later in the text) is also derived using simple
statistical arguments.  Finally, we have a brief discussion in Section~IV.

\section{The Burst Model}
Consider a two-dimensional square lattice with periodic boundary conditions.
(Rules for other lattice geometries are similar.)
We denote the energy stored at each lattice site ${\bf r}$ by $E({\bf r})
(\ge 0)$.  At each timestep, the system is parallel updated
according to the following rules \cite{Chan95}:
\medskip
\begin{enumerate}
\item Energy introduction: a unit of energy is added to a randomly chosen site
 ${\bf r_0}$, i.e.
$E({\bf r_0}) \rightarrow E({\bf r_0})+1$.
\item Fire triggering: if the energy in the chosen site is greater than or
equals to a fix threshold $E_{c1}$, the energy
will be dissipated (or burnt), i.e. if $E({\bf r_0}) \ge E_{c1}$, then
$E({\bf r_0}) \rightarrow 0$.
\item Fire propagation: if the energy in the nearest neighbor of a burning site
is greater than or equal to another fix threshold $E_{c2}(\leq E_{c1})$, the
energy in that neighboring site will be also dissipated. That is, if
$E({\bf r'_0}) \ge E_{c2}$, then $E({\bf r'_0}) \rightarrow 0$, where
${\bf r'_0}$ are the nearest neighbors of the burning site ${\bf r}$. This
process is repeated until the fire cannot propagate any longer. And we also
assume that fire propagation is so fast that it takes only one timestep to
set the relevant sites on fire.
\item Material diffusion: diffusion takes place at every site, i.e. $E({\bf r})
\rightarrow E({\bf r})+\Delta E({\bf r})$ for all ${\bf r}$, where
$\Delta E$ depends on the geometry of the lattice and the diffusion constant
$D$\footnote{Nevertheless, our study is focused on diffusionless system in
this paper.}.
\end{enumerate}
\medskip
The system is allowed to evolve for a sufficiently long time in order to attain
a steady state before any statistics is taken.

The behavior of the burst model depends on four external
parameters, namely, $E_{c1}$, $E_{c2}$, $D$, and the size of each energy blob
in comparison with the size of a lattice cell.  Depending on various choice of
these parameters, the occurrence probability $P(S)$ of dissipation size $S$
falls into one of the following three types of behaviors \cite{Chan95}:
(a) $P(S)$ shows a localized peak which is a signature of a super-critical
state; (b) $P(S)$ shows an early exponential cut-off which can be identified as
a sub-critical state; and (c) power law is observed in $P(S)$ at the transition
point between the super-critical and sub-critical phase.

\section{Statistics In The Super-critical Phase}
Let us introduce a few concepts before discussing the super-critical phase
statistics and the relationship between the burst model at super-critical phase
and percolation.  A site is called {\em burnable} if its energy content is
greater than or equals to $E_{c2}$.  Thus, a burnable site will catch fire when
any one of its nearest neighbors burns.  Then, we define a {\em burnable
cluster} to be a collection of maximal\footnote{The collection of sites is
maximal in the sense of inclusion.} burnable sites such that all of them catch
fire in the next timestep provided that any one of them is on fire
\cite{Chan95}.  The size of a burnable cluster $s$ is defined as the number of
sites it contains.

During the evolution of the system, clusters of sites grow continuously while
they burn suddenly to empty spaces (or voids).  And it is interesting to note
that the dissipation size $S$ (i.e., the total energy released in the burning
of a cluster) is generally proportional to the cluster size.  The behavior of
$P(S)$ under different choice of parameters can be understood in terms of the
size of the burnable clusters \cite{Chan95}.  In particular, if there is only
a few large burnable clusters in the system immediately before most of the
burnings, $P(S)$ will show a peak indicating that majority of the energy
release in the system occurs in the ``large events'' which extend across the
whole system.  Therefore, the system is in the super-critical phase.  Finally,
we define the {\em coverage fraction} $\rho$ to be the percentage of sites in
the system that found to be burnable.

As we can see in Fig.~\ref{snap}, which is a snapshot of a typical system
immediately after a large energy dissipation, a large void is observed with
cluster islands of different sizes lying inside.  So, in general, a burnable
cluster for the system (in two-dimension) is not compact.

To investigate the relation between our model and percolation, we consider a
region of space with linear size $L$ inside the system.  If the typical length
of a burnable cluster $\xi$ is much less than $L$, we expect most of the
dissipation in this area is initiated and confined or localized inside the
region.  On the other hand, if $\xi \gtrsim L$, the coverage fraction in this
area would gradually grow to a value $\rho_b$ before a large dissipation
sweeps through the whole region.  Therefore, if the system is super-critical,
we can take a sufficiently large $L$ and still have non-negligible probability
of finding burnable clusters with size $\agt L^2$.  In order to burn down such
a large cluster, the coverage fraction around in the large burnable cluster
immediately before it burns must be greater than or equal to the (site)
percolation threshold $\rho_{c}$ of the corresponding lattice.  In general,
the (time) mean coverage fraction $\bar{\rho}$ depends on the large as well as
small dissipations.  But at the super-critical phase, large dissipations
dominate the behavior of the system and hence we can safely neglect the
effect of small dissipations.

Using the same argument, we know that the coverage fraction $\rho$ attains its
local minimum value just after a fire.  Then as energy is gradually introduced
into the system, $\rho$ increases.  However, the rate of increase $\dot{\rho}$
will, in general, flatten out gradually with time due to the fact that adding
energy into a site which is already burnable does not increase $\rho$.
Eventually, $\rho$ reaches its local maximum value immediately before a fire.
A typical time evolution of $\rho$ for a super-critical system is plotted in
Fig.~\ref{F:2}.  In this respect, the time evolution of burnable clusters in
the
burst model can be viewed as a dynamical percolation problem where the
``conductance probability'' gradually increases with time.  Any sudden drop in
the conductance probability is due to a fire.  Finally, we like to remark that
both $\rho_b$ and $\bar{\rho}$ are functions of $E_{c1}$ and $E_{c2}$.

\subsection{The Case When $\protect\bbox{E_{c2} = 1}$}
We first consider an easy case with $D = 0$, $E_{c2} = 1$ and $E_{c1} \gg
E_{c2}$.  With this set of parameters, the system is in the super-critical
phase \cite{Chan95}.  Since $E_{c2} = 1$, an empty site turns burnable once an
energy unit is dropped onto it.  Moreover, burning is the only way to turn a
burnable site back to an empty site because $D = 0$.  Thus, a burnable site is
formed out of an empty site with probability $(1-\rho)/N$ at each timestep
where $N$ is the total number of sites in the system.  That is,
\begin{equation}
 \frac{d\rho}{dt} = k(1-\rho) \label{E:1}
\end{equation}
for some positive constant $k$. Integrating Eq.~(\ref{E:1}) gives us
\begin{equation}
 \rho (t) = 1 - e^{-kt} \mbox{~,} \label{E:2}
\end{equation}
where $t$ is the time since the site is last burnt (and hence $\rho (0)
\approx 0$).  Since the system is in the super-critical phase and $E_{c2} = 1$,
it is almost sure that immediately before each burning, almost the entire
system is covered by a single large burnable cluster.  So after each burning,
almost all the sites becomes void.  Everything starts over again almost
independent of the history of the system.  Thus, we may relate the mean
coverage fraction, $\bar{\rho}$, to the average coverage fraction attained just
before burning, $\rho_b$, by
\begin{equation}
 \bar{\rho}  = \left< 1 + \frac{\rho'}{\ln(1-\rho')} \right> \approx
 1+\frac{\rho_b}{\ln(1-\rho_b)} \mbox{~,} \label{E:3}
\end{equation}
where the expectation is taken over all the values of coverage fractions,
$\rho'$, just before burnings.

In the $E_{c1} \longrightarrow \infty$ limit (while the lattice size remains
fix), every grid point the system must be covered by a single burnable cluster
right before each fire.  So $\rho_b = \bar{\rho} = 1$, which agrees with the
prediction of Eq.~(\ref{E:3}).  As we lower the value of $E_{c1}$, some sites
may not be contained in the burnable cluster before a fire.  Therefore, we
expect both $\rho_b$ and $\bar{\rho}$ to be less than 1.  At this point,
Eq.~(\ref{E:3}) still holds.  As we further decrease the value of $E_{c1}$,
the system eventually enters the critical, and then follows by the sub-critical
phase.  As there is negligible fire which extends system-wide, fluctuation of
$\rho$ about its (time) mean value $\bar{\rho}$ is small.  So for a infinitely
large system, $\rho_b$ equals $\bar{\rho}$ and their relation given by
Eq.~(\ref{E:3}) is no longer valid.  Finally, as $E_{c1} \longrightarrow
E_{c2}$, both $\rho_b$ and $\bar{\rho}$ goes to 0.

In order to study the local properties of the system near a burnable cluster,
we define $\left( \rho_b \right)_{\text{Local}}$ to be the (time) average
coverage fraction immediately before a fire at a sufficiently small region
centered at the triggering site.  Using the same argument in the derivation of
Eq.~(\ref{E:3}), we find that
\begin{equation}
 \bar{\rho} \approx  1 - \frac{\left( \rho_b \right)_{\text{Local}} - \rho
 (0)}{\log \left( 1-\rho (0) \right) - \log \left( 1 - \left( \rho_b
 \right)_{\text{Local}} \right)} \mbox{~,} \label{E:3a}
\end{equation}
where $\rho (0)$ is determined by the amount of energy remains after the last
fire in the nearby region, which should be close to 0.  We expect $\left(
\rho_b \right)_{\text{Local}}$ to be greater than its corresponding global
value $\rho_b$.  While $\rho_b$ does not fluctuate in the limit of large system
size in the sub-critical phase, $\left( \rho_b \right)_{\text{Local}}$ near
a burning cluster can deviate from $\rho_b$.

The above discussions are verified by numerical simulations\footnote{All
computer simulations in this paper are done in SUN SPARC~10 workstations in
University of Hong Kong, and SGI Challenge in IAS}.  We fix $E_{c2} = 1$ and
$D = 0$ on a $512\times 512$ square lattice.  $\bar{\rho}$ is obtained by
calculating the time mean of $\rho$.  On the other hand, $\rho_b$ and $\left(
\rho_b \right)_{\text{Local}}$ are obtained by calculating the median of their
corresponding values immediately before each fire.  Their values obtain this
way will not be seriously affected by the ``pre-mature'' burnings.  As shown in
Table~\ref{T:a}, Eq.~(\ref{E:3}) agrees very well with our numerical experiment
in the super-critical phase. In addition, the relation between $\bar{\rho}$ and
$\left( \rho_b \right)_{\text{Local}}$ in both the super-critical and
sub-critical phase is also verified.  As one can notice from Table~\ref{T:a},
the value of $\bar{\rho}$ predicted from $\left( \rho_b \right)_{\text{Local}}$
deviates systematically from the observed valued of $\bar{\rho}$ in the
sub-critical phase.  The deviation can be explained as follows:
Eq.~(\ref{E:3a}) tells us that the predicted value of $\bar{\rho}$ increases
when we increase $\rho (0)$ but keeping $\rho_b$ fixed.  Consequently, when
$\bar{\rho}$ is fixed, $\rho_b$ decreases with increasing $\rho (0)$.  Thus, in
the sub-critical phase whose $\rho (0) > 0$, the prediction of $\bar{\rho}$ in
Table~\ref{T:a} using the assumption that $\rho (0) = 0$ is systematically
drifted to the lower side.

All the above observations are also verified for honeycomb, triangular, and
simple cubic lattices (see Table~\ref{T:a}).  In the supercritical phase,
Eq.~(\ref{E:3}) holds very well in all these lattices.

Since we only use a modest set of assumptions in deriving these relations, we
expect Eqs.~(\ref{E:3})-(\ref{E:3a}) to hold for a number of models with
similar growth mechanisms as long as there is a separation of timescale between
the gradual growth and sudden dissipation.  To test this hypothesis, we look
into the version of FFM proposed by Drossel {\em et al.} \cite{FFM}.  We have
performed numerical experiment on a $1024\times 1024$ square lattice with the
probability of tree grow and lightning per site per timestep being $1\times
10^{-9}$ and $1\times 10^{-13}$ respectively.  The average coverage fraction
$\bar{\rho}$ is found to be 0.408(1) which is consistent with the value
0.4081(7) found recently by Clar {\em et al.} \cite{Dross94}.  We also obtain
$\rho_b = 0.423(2)$ and $\left( \rho_b \right)_{\text{Local}} = 0.62(3)$.  If
we assume that $\rho (0) = 0$, $\bar{\rho}$ deduced from the local coverage
fraction equals 0.359(2) which is again drifted to the lower side.
Nevertheless, within an error of 10\% or so, Eq.~(\ref{E:3a}) gives a
reasonably good estimate of $\bar{\rho}$ based on $\left( \rho_b
\right)_{\text{Local}}$.  In this respect, we believe that Eq.~(\ref{E:3a}) is
a good approximation for models with growth mechanisms similar to the burst
model or the FFM.

\subsection{The Case When $\protect\bbox{E_{c2} > 1}$}
The situation is more complicated when the system is in super-critical phase
with $E_{c2}>1$.  For example, if we take $E_{c1} = 16$, $E_{c2} = 2$ and
$D = 0$ in the $512\times 512$ square lattice, Table~\ref{T:b} tells us that
$\rho_b = 0.874(3)$ and $\bar{\rho} = 0.502(1)$.  Although the system is in
the super-critical phase, Eq.~(\ref{E:3}) does not hold.  This discrepancy is
due to the invalidity of Eq.~(\ref{E:1}).  Since the adding an unit of energy
to a site with $E < E_{c2}$ does not necessarily turn it into a burnable site,
we expect $d\rho / dt$ to be smaller than $k (1-\rho )$.

For simplicity, we consider a diffusionless system and write $E_{c2} = n$.
We also define $\rho_j$ to be the coverage fraction of the system by sites
with energy equals $j$ for $j = 0,1,2,\ldots ,n-1$.  Moreover, $\rho_n$ is the
coverage fraction of the system by sites with energy greater than or equals to
$n$ (hence they are burnable and $\rho_n \equiv \rho$).  It is easy to see
that
\begin{mathletters}
\begin{equation}
 \frac{d\rho_0}{dt} = - k \rho_0 \mbox{~,}
\end{equation}
\begin{equation}
 \frac{d\rho_j}{dt} = k \rho_{j-1} - k \rho_{j} \hspace{0.25in} \mbox{for~} j
 = 1,2,\ldots ,n-1,
\end{equation}
and
\begin{equation}
 \frac{d\rho_n}{dt} = k \rho_{n-1} \mbox{~,}
\end{equation}
\label{E:4}
\end{mathletters}
\par\noindent
where $k$ is a positive constant.  Note that the coverage fractions satisfy the
conversation law
\begin{equation}
 \sum_{j=0}^{n} \rho_j = 1 \mbox{~.} \label{E:5}
\end{equation}
Clearly, Eq.~(\ref{E:4}c) reduces to Eq.~(\ref{E:1}) when $n = 2$.

Given the initial coverage fractions, we can integrate Eqs.~(\ref{E:4}) one
at a time, which can eventually calculate the time evolution of $\rho_j$ for
all $j < n$.  Then $\rho_n (t)$ can be computed using Eq.~(\ref{E:5}).
After some simple but long computation, the solutions of Eqs.~(\ref{E:4}) are
given by
\begin{mathletters}
\begin{equation}
 \rho_j (t) = \sum_{m = 0}^{j} \frac{ \left( k t \right)^m}{m!} \rho_{j-m} (0)
 e^{-kt} \hspace{0.25in} \mbox{for~} j = 0,1,2,\dots ,n-1,
\end{equation}
and
\begin{equation}
 \rho_n (t) = 1 - \sum_{j=0}^{n-1} \sum_{m=0}^{j} \frac{ \left( k t
 \right)^m}{m!} \rho_{j-m} (0) e^{-kt} \mbox{~,}
\end{equation}
\label{E:6}
\end{mathletters}
\par\noindent
where $t$ is the time since the last burning.  Just after a large energy
dissipation, $\rho_0 (0) \approx \rho_b$, $\rho_j (0) \approx ( 1 - \rho_b ) /
(n - 1)$ for $j=1,2,\ldots ,n-1$, and $\rho_n (0) = 0$ are reasonable
approximations.  Thus, Eq.~(\ref{E:6}b) becomes
\begin{equation}
 \rho_n (t) \approx 1 - \rho_b \sum_{j=0}^{n-1} \frac{\left( kt \right)^j}{j!}
 e^{-kt} - \frac{1-\rho_b}{n-1} \sum_{j=1}^{n-1} \frac{\left( kt \right)^m}{m!}
 (n-2-j) e^{-kt} \mbox{~.} \label{E:7full}
\end{equation}
In particular, if $E_{c1}$ is very large, $\rho_b \approx 1$, and
Eq.~(\ref{E:7full}) becomes
\begin{equation}
 \rho_n (t) \approx 1 - e^{-kt} \sum_{m=0}^{n-1} \frac{ \left( kt\right)^m}{m!}
 \mbox{~.} \label{E:7}
\end{equation}
\indent
Unfortunately, there are no simple close forms for the time $T$ when $\rho (T)
\equiv \rho_n (T) = \rho_b$ in general for both Eqs.~(\ref{E:7full})
and~(\ref{E:7}). So we do not have a simple relationship between $\bar{\rho}$
and $\rho_b$ as in the case when $E_{c2} \equiv n = 1$.

To verify the validity of Eq.~(\ref{E:7full}), we numerically solve for $kT$
when $\rho_n (T) = \rho_b$ and $n = 2$.  Once $kT$ is found, we can numerically
integrate Eq.~(\ref{E:7full}) to obtain $\bar{\rho}$.  The results are
tabulated
in Table~\ref{T:b}.  As we can see, the prediction of Eq.~(\ref{E:7full})
agrees
quite well with our numerical experiments.

\section{Discussions}
In summary, we have investigated the statistics of the burst model at its
super-critical phase.  We have a simple picture for the time evolution of the
coverage fraction $\rho$.  Using simple statistical argument, we also derive
a relationship between the (time) mean coverage fraction $\bar{\rho}$ and the
average coverage fraction just before a fire $\rho_b$.  The relation agrees
quite well with our numerical simulations within an error level of a few
percent.  Moreover, we find a local version of this relationship (see
Eqs.~(\ref{E:3})-(\ref{E:3a})).  Since these two relations are derived using a
modest set of assumptions, we expect that they are valid in many other growth
models with dissipations.  Our preliminary analysis on the FFM tells us that
this is indeed the case.  And it will be very interesting to apply the same
analysis to both the burst model in higher spatial dimensions and other growth
and dissipation models.

\acknowledgements
This work is supported in part by the DOE grant DE-FG02-90ER40542.

\begin{table}
 \caption{Values of the average coverage fractions when $D = 0$ and $E_{c2} =
1$
  in various underlying lattices with different values of $E_{c1}$.}
 \vspace{0.35in}
 \squeezetable
 \begin{tabular}{c|cddddd|c}
  Lattice & $E_{c1}$ & $\bar{\rho}$ & $\rho_b$ & $\left( \rho_b
   \right)_{\text{Local}}$ & $\left( \bar{\rho}
   \right)_{\text{proj1}}$\tablenotemark[1] & $\left( \bar{\rho}
   \right)_{\text{proj2}}$\tablenotemark[2] & Phase \\ \hline
  $512\times 512$ & 4 & 0.334(2) & 0.338(4) & 0.55(4) & & 0.31(2) & \\
   \cline{2-7}
  Square & 6 & 0.381(1) & 0.382(2) & 0.63(3) & & 0.36(2) & Sub-critical \\
   \cline{2-7}
  & 8 & 0.400(4) & 0.41(1) & 0.66(1) & & 0.388(7) & $\uparrow$ \\
   \cline{2-7}
  & 10 & 0.426(5) & 0.61(1) & 0.71(1) & 0.35(1) & 0.426(8) &
   $\downarrow$ \\ \cline{2-7}
  & 11 & 0.509(5) & 0.79(1) & 0.800(4) & 0.49(1) & 0.503(4) &
   Super-critical \\ \cline{2-7}
  & 12 & 0.580(4) & 0.872(5) & 0.872(4) & 0.576(6) & 0.576(5) & \\
   \cline{2-7}
  & 13 & 0.633(3) & 0.920(3) & 0.920(3) & 0.636(4) & 0.636(4) & \\
   \hline
  $512\times 512$ & 8 & 0.473(2) & 0.476(2) & 0.73(4) &  & 0.44(3) &
   Sub-critical \\ \cline{2-7}
  Honeycomb & 10 & 0.495(4) & 0.52(1) & 0.76(3) & & 0.47(3) & $\uparrow$ \\
   \cline{2-7}
  & 12 & 0.530(3) & 0.81(2) & 0.82(1) & 0.51(2) & 0.52(1) & $\downarrow$ \\
   \cline{2-7}
  & 14 & 0.666(1) & 0.942(2) & 0.941(1) & 0.669(3) & 0.668(2) & super-critical
   \\ \hline
  $512\times 512$ & 6 & 0.323(4) & 0.326(4) & 0.52(4) & & 0.29(3) &
   sub-critical \\ \cline{2-7}
  Triangular & 8 & 0.337(3) & 0.400(5) & 0.54(3) & & 0.30(3) & $\uparrow$ \\
   \cline{2-7}
  & 10 & 0.465(1) & 0.753(4) & 0.754(4) & 0.462(3) & 0.462(3) & $\downarrow$ \\
   \cline{2-7}
  & 12 & 0.586(1) & 0.880(2) & 0.880(3) & 0.584(2) & 0.584(3) & super-critical
   \\ \hline
  $64\times 64\times 64$ & 6 & 0.220(2) & 0.230(6) & 0.30(3) & & 0.15(2) &
   sub-critical \\ \cline{2-7}
  Simple Cubic & 8 & 0.294(4) & 0.488(4) & 0.510(5) & 0.271(3) & 0.285(4) &
   $\updownarrow$ \\ \cline{2-7}
  & 10 & 0.464(2) & 0.756(2) & 0.756(4) & 0.464(2) & 0.464(3) & super-critical
 \end{tabular}
 \tablenotetext[1]{$\left( \bar{\rho} \right)_{\text{proj1}}$ is the value of
  $\bar{\rho}$ obtained as if Eq.~(\protect\ref{E:3}) is correct.  The
  difference between this value and $\bar{\rho}$ we measure (in column~3) tells
  us how good Eq.~(\protect\ref{E:3}) is.  We only show this prediction when
the
  system is in super-critical state.}
 \tablenotetext[2]{$\left( \bar{\rho} \right)_{\text{proj2}}$ is the value of
  $\bar{\rho}$ obtained by substituting $\left( \rho_b \right)_{\text{Local}}$
  into Eq.~(\protect\ref{E:3a}) and assuming $\rho (0) = 0$.}
 \label{T:a}
\end{table}
\newpage
\begin{table}
 \caption{Values of the average coverage fractions when $D = 0$ and $E_{c2} =
  2$ in $512\times 512$ square lattice using different values of $E_{c1}$.}
 \vspace{0.35in}
 \begin{tabular}{cddddd|c}
  $E_{c1}$ & $\bar{\rho}$ & $\rho_b$ & $\left( \rho_b \right)_{\text{Local}}$
   & $\left( \bar{\rho} \right)_{\text{proj1}}$\tablenotemark[1] & $\left(
   \bar{\rho} \right)_{\text{proj2}}$\tablenotemark[2] & Phase \\ \hline
  8 & 0.338(1) & 0.338(1) & 0.60(5) & & 0.32(4) & sub-critical \\ \cline{1-6}
  10 & 0.358(2) & 0.361(3) & 0.64(4) & & 0.34(4) & $\uparrow$ \\ \cline{1-6}
  12 & 0.368(2) & 0.387(3) & 0.68(1) & & 0.365(8) & $\downarrow$ \\ \cline{1-6}
  14 & 0.379(2) & 0.557(3) & 0.685(6) & 0.297(3) & 0.368(5) & super-critical \\
   \cline{1-6}
  16 & 0.502(1) & 0.874(3) & 0.875(5) & 0.503(3) & 0.504(5) &
 \end{tabular}
 \tablenotetext[1]{$\left( \bar{\rho} \right)_{\text{proj1}}$ is the value of
  $\bar{\rho}$ obtained as if Eq.~(\protect\ref{E:7full}) is correct.  The
  difference between this value and $\bar{\rho}$ we measure (in column~2) tells
  us how good Eq.~(\protect\ref{E:7full}) is.  We only show this prediction
when
  the system is in super-critical state.}
 \tablenotetext[2]{$\left( \bar{\rho} \right)_{\text{proj2}}$ is the value of
  $\bar{\rho}$ obtained using the local version of
Eq.~(\protect\ref{E:7full}).}
 \label{T:b}
\end{table}

\begin{figure}
 \caption{Gray scale snapshot of a $512\times 512$ system with $E_{c1} = 8$,
  $E_{c2} = 1$ and $D = 0$ just after a large dissipation showing a large void
  inside which there are some small burnable clusters.}
 \label{snap}
\end{figure}

\begin{figure}
 \caption{The time evolution of the coverage fraction $\rho$ (in solid line) in
  a typical super-critical system.  Here, we use $E_{c1} = 13$, $E_{c2} = 1$,
  and $D = 0$ in a $512\times 512$ square lattice.  The values of $\bar{\rho}$
  and $\rho_b$ are shown (in dash lines) for comparison.}
 \label{F:2}
\end{figure}
\end{document}